\PassOptionsToPackage{numbers,sort&compress}{natbib}
\documentclass{wiley-article}

\usepackage{graphicx}
\graphicspath{{figures/}}
\usepackage[space]{grffile}
\usepackage{latexsym}
\usepackage{textcomp}
\usepackage{longtable}
\usepackage{tabulary}
\usepackage{booktabs,array,multirow}
\usepackage{amsfonts,amsmath,amssymb}
\usepackage{url}
\usepackage{hyperref}
\hypersetup{colorlinks=false,pdfborder={0 0 0}}
\usepackage{etoolbox}
\makeatletter
\patchcmd\@combinedblfloats{\box\@outputbox}{\unvbox\@outputbox}{}{}
\makeatother
\newif\iflatexml\latexmlfalse
\AtBeginDocument{\DeclareGraphicsExtensions{.pdf,.PDF,.eps,.EPS,.png,.PNG,.tif,.TIF,.jpg,.JPG,.jpeg,.JPEG}}

\usepackage[utf8]{inputenc}
\usepackage[english]{babel}
\usepackage[version=4]{mhchem}
\usepackage{braket}
\usepackage{tabularx}
\renewcommand{\arraystretch}{1.15}

\papertype{Original Article}
\title{Seniority-zero Quadratic Canonical Transformation Theory}

\author[1]{Daniel F. Calero-Osorio}
\author[1]{Paul W. Ayers}
\affil[1]{Department of Chemistry, McMaster University, Hamilton, Ontario, L8S 4M1, Canada}

\corraddress{Paul W. Ayers, Department of Chemistry, McMaster University, Hamilton, Ontario, L8S 4M1, Canada}
\corremail{ayers@mcmaster.ca}
\runningauthor{Calero-Osorio and Ayers}

\begin{document}
\maketitle
\selectlanguage{english}
\begin{abstract}
We propose a method to solve the Schrödinger equation for systems with static/strong electron correlation using Hamiltonian transformations. Building on our previous work on seniority-zero canonical transformation theory \cite{10.1063/5.0309818}, which seeks a unitary transformation that maps the Hamiltonian into the seniority-zero space, this method presents an alternative way of evaluating the Baker--Campbell--Hausdorff (BCH) expansion based on quadratic canonical transformation theory. The extension aims to relax the small-generator constraint by allowing approximate four-body contributions in the expansion, thus expanding the class of excitations previously allowed in SZ-LCT, where only approximate three-body operators were retained. Numerical tests reveal that the seniority-zero quadratic canonical transformation method (SZ-QCT) delivers good accuracy, with most errors within chemical accuracy. In particular, SZ-QCT shows sub-millihartree errors in cases where larger generators are needed to recover the residual dynamic correlation. The computational scaling of SZ-QCT is the same as that of SZ-LCT, $\mathcal{O}(N^8/n_c)$, where $n_c$ is the number of cores available for the computation.

\end{abstract}

\section{Introduction}
Electron correlation is usually divided into two components, dynamic (associated with the correlated motion of electrons due to their Coulombic repulsion) and static (associated with nearly degenerate electron configurations). Dynamic correlation effects are incorporated by adding excited determinants to the reference wave function; typically the dominant effects come from single- and double- excitations of the reference. When static correlation is neglible, we say the system is weakly correlated. In such cases, the reference is usually a single Slater determinant (e.g., the Hartree-Fock determinant), and the corrections are introduced through configuration interaction (CI) \cite{doi:https://doi.org/10.1002/9781119019572.ch11,szabo1996modernCI,Lowdin1955,KNOWLES1984315,OlsenRoos1988,SherrillSchaefer1999,ShavittBartlett2009,SzalayMuller2012}, coupled cluster (CC) \cite{ShavittBartlett2009,doi:https://doi.org/10.1002/9781119019572.ch13,szabo1996modernCC,Coester1960,Cizek1966,paldus1982relationship,CrawfordSchaefer2000,BartlettMusial2007}, or many-body perturbation theory (MBPT) \cite{ShavittBartlett2009,PhysRev.46.618,doi:https://doi.org/10.1002/9781119019572.ch14,szabo1996modernMBPT,goldstone1957derivation,Bloch1958,FetterWalecka1971,LindgrenMorrison1985}. When static correlation is nonnegligible, one needs a reference wavefunction that includes multiple electron configurations and---in the limit of strong correlation---perhaps even a combinatoric number electron configurations. 
Common static correlation methods include complete active space self-consistent field (CASSCF) \cite{ROOS1980157,siegbahn1981complete,roos1987complete,schmidt1998construction,szalay2012multiconfiguration} and complete active space configuration interaction (CASCI) \cite{slavivcek2010ab,szalay2012multiconfiguration,levine2021cas,fales2017complete}, where all electron configurations in an ``active space'' of (valence) orbitals are included in the reference. While the distinction between dynamic and static correlation is somewhat arbitrary, it is convenient to treat static and dynamic correlation separately. Typically this is done by first treating the static (multiconfigurational) correlation and subsequently adding corrections for dynamic correlation, though dynamic-then-static \cite{liuIdeasRelativisticQuantum2010a,durandDirectDeterminationEffective1983,lindgrenCoupledclusterApproachManybody1978,liuPASPT2SizeExtensiveSizeConsistent2026} and static-dynamic-static methods are emerging \cite{huangIVIIterativeVector2017,leiFurtherDevelopmentSDSPT22017,liuICIIterativeCI2016a,
liuSDSStaticdynamicstaticFramework2014a,zhangFurtherDevelopmentICIPT22021,leiSDSPT2sSDSPT2Selection2025}. However, we use the conventional static-then-dynamic approach in this work.

Static-then-dynamic approaches typically use the multireference extensions of single-reference dynamic correlation methods like multireference M\o ller--Plesset perturbation theory (MRMP) \cite{HIRAO1992374,Hirao1993,WolinskiPulay,Wolinski,Nakano1993,GrimmeWaletzke2000,witek2002intruder}, second-order complete active space perturbation theory (CASPT2) \cite{Roos1982,doi:10.1021/j100377a012,Andersson1992,Andersson1995,Forsberg1997,Finley1998,Ghigo2004}, and multireference configuration interaction (MRCI) \cite{BuenkerPeyerimhoff1974,10.1063/1.455556,10.1063/1.439365}. In these methods, the primary working object is the wave function, and dynamic correlation effects are incorporated through either perturbative corrections or expansions in excited determinants.

Alternative approaches use the Hamiltonian as the principal working object. Examples include multireference coupled cluster (MRCC) \cite{JeziorskiMonkhorst1981,RittbyBartlett1991,PaldusPiecuchPylypowJeziorski1993,mahapatra1999size,lyakh2012multireference,kohn2013state}, canonical transformation (CT) theory \cite{10.1063/1.3086932,10.1063/1.2196410,C2CP23767A,Neuscamman01042010}, and the driven similarity renormalization group (DSRG) \cite{10.1063/1.4890660,doi:10.1021/acs.jctc.5b00134,10.1063/1.4947218,10.1063/5.0059362}. In general, these methods seek a similarity transformation that downfolds the Hamiltonian into an active space, and the transformation can be unitary (e.g., CT or DSRG) or nonunitary (e.g., MRCC). In either case, the transformation is generally evaluated using the Baker-Campbell-Hausdorff (BCH) expansion. Unlike in single-reference CC, the BCH expansion does not truncate naturally for multireference methods, so approximation stategies must be designed, implemented, and tested.

For MRCC approaches, early formulations either restricted the operator manifold to (approximately) commuting subsets or truncated the BCH expansion at an order chosen to avoid terms involving nested commutators of noncommuting cluster operators \cite{https://doi.org/10.1002/qua.560190203,BANERJEE1983297,HOFFMANN1987451,10.1063/1.454125}. More recent developments discard purely active excitations, which leads to a truncation at the eightfold commutator in the amplitude equations when the cluster operator contains up to two-body excitations \cite{10.1063/1.3559149,10.1063/1.3592786}. Even with this truncation, evaluating up to eight commutators remains computationally infeasible, so an additional approximation---retaining only the first two commutators in the amplitude equations---is typically imposed.\cite{evangelistaOrbitalinvariantInternallyContracted2011,
feldmannRenormalizedInternallyContracted2024a,
hanauerPilotApplicationsInternally2011,
szenesEfficientImplementationSpinFree2026}

On the other hand, unitary-transformation methods such as CT theory and DSRG have an inherently nonterminating BCH series because the transformation generator is anti-Hermitian. For that reason, the recursive commutator approximation (RCA) was introduced in CT theory and later adopted by DSRG. This approximation allows one to rewrite the commutator between two operators as a series of products of one- and two-body operators and reduced density matrices (RDMs). By truncating each commutator at the one- and two-body level, the BCH expansion can be evaluated approximately and recursively to a chosen order. The RCA uses an operator decomposition (OD) formula to approximate three-body interactions as superpositions of one- and two-body interactions, while higher-body interactions are neglected. An extension of this approximation later included approximate four-body contributions in the BCH expansion, although the resulting improvements were largely limited to single-determinant references \cite{10.1063/1.3086932}.

Many state-specific MRCC and DSRG formulations begin by partially including static correlation through the direct selection of important electron configurations (multireference and multiconfigurational references) or through the direct selection of a (generalized) active orbital space (complete, generalized, or restricted active space approaches) \cite{doi:https://doi.org/10.1002/9780470142943.ch7,malmqvistRestrictedActiveSpace1990,roosMulticonfigurationalQuantumChemistry2016,maGeneralizedActiveSpace2011}. These selection techniques are problematic, as they require physical insight or \textit{ad hoc} heuristics; moreover, the important configurations or orbitals usually differ across different regions of a molecular potential energy surface \cite{kaufoldAutomatedActiveSpace2023,kellerSelectionActiveSpaces2015,kingRankedOrbitalApproachSelect2021,sayfutyarovaAutomatedConstructionMolecular2017,steinAutomatedSelectionActive2016,tothComparisonMethodsActive2020}. In state-specific MRCC and perturbative DSRG approaches such as DSRG-MRPT2, dynamic correlation is then incorporated through the Hamiltonian downfolding process, and the remaining static correlation is recovered through orbital relaxation, placing these methods in the static-dynamic-static family. The complications associated with selecting the relevant orbitals in the initial static-correlation step motivated us, and others, to consider methods based on seniority-zero (doubly occupied) reference states \cite{weinholdReducedDensityMatrices1967,veillardCompleteMulticonfigurationSelfconsistent1967,cookDoublyoccupiedOrbitalMCSCF1975,carboGeneralMulticonfigurationalPaired1977,bytautasSeniorityNumberDescription2015,alcobaHybridConfigurationInteraction2014a,CaleroOsorio2025SZ}. In such cases, no active orbitals are selected; instead, one considers \textit{all} electron configurations in which every spatial orbital is either doubly occupied or empty. Mathematically, this approximation is appealing because seniority-zero configuration interaction and maximum-seniority configuration interaction are the only size-consistent selected configuration interaction methods. Practically, seniority-zero reference states have been observed to provide a qualitatively correct description of many chemical-bond-breaking processes. Remarkably, all eigenstates of seniority-zero Hamiltonians can be expressed as product wave functions \cite{10.1063/5.0296924}, specifically number-symmetry-broken antisymmetric products of interacting geminals \cite{silver1970bilinear,silver1969natural,johnsonStrategiesExtendingGeminalbased2017b}.

In previous work \cite{10.1063/5.0309818}, we introduced the seniority-zero linear canonical transformation method (SZ-LCT), which seeks to downfold the molecular Hamiltonian into the seniority-zero subspace, where seniority-zero methods are \textit{exact}. This is achieved through a unitary transformation, by evaluating the BCH expansion using the RCA and minimizing the non-seniority-zero couplings of the transformed Hamiltonian. This method showed high accuracy for ground-state energy calculations, with errors on the order of $10^{-4}E_h$. However, the method showed a strong dependence on the small-generator constraint inherent to the RCA, causing significant energy deviations when the reference wave function was not an accurate description of the system. To reduce the truncation error,  we designed the late-truncated seniority-zero canonical transformation method (LT-SZCT) \cite{doi:10.1021/acs.jctc.5c01892}, designed to improve the accuracy of SZ-LCT by evaluating the first two commutators of the BCH expansion exactly and using the RCA for the remaining terms. In this way, the method achieved higher accuracy and stability while maintaining the same formal computational scaling. However, unlike SZ-LCT, LT-SZCT is state-specific in nature, as its objective function targets the ground-state energy rather than the Hamiltonian couplings.

Here, we introduce the seniority-zero quadratic canonical transformation method, a direct extension of SZ-LCT in which the commutator approximation is extended to partially retain four-body contributions in the BCH series. In this way, the constraint on the generator norm is relaxed, and larger generators can be included without breaking the RCA.

The remainder of this paper is organized as follows. Sections \ref{spinfree} and \ref{SZ-QCT} review the requisite material on spin-free second-quantized operators and canonical transformation theory, respectively. Sec. \ref{sz-reference} explains our choice of the seniority-zero reference wave function. The commutator approximation and the recursive formula for the BCH expansion is given in Sec.~\ref{commutator_approx}. Then, we present the operator decomposition process in Sec.~\ref{commutator_approx}. Details on our computational implementation are given in section \ref{implementation}. In Sec. \ref{results} we apply the method to three molecules: \ce{H_6}, \ce{BeH2}, and \ce{BH}. Finally, Sec. \ref{conclusions} offers our conclusions and outlines prospects for future work. 
 
\section{Theory}

\subsection{Spin-free operators}
\label{spinfree}
It is useful to employ spin-free operators because they ensure the theory is full spin-adapted and, moreover, they (vastly) reduce the number of terms that need to be treated. Recall that spin-free operators are defined as excitation operators where all the spin degrees of freedom are summed over:
\begin{equation}
\begin{aligned}
    \hat{E}^{p_1}_{q_1}&=\sum_{\sigma= \alpha, \beta}\hat{c}^{\dagger}_{p_1 \sigma}\hat{c}_{q_1 \sigma}, \\[8pt]
    \hat{E}^{p_1 p_2}_{q_1 q_2}=&\sum_{\sigma,\tau= \alpha, \beta}\hat{c}^{\dagger}_{p_1 \sigma}\hat{c}^{\dagger}_{p_2 \tau}\hat{c}_{q_2 \tau}\hat{c}_{q_1 \sigma},\\[8pt]
    \hat{E}^{p_1 p_2 p_3}_{q_1 q_2 q_3}=&\sum_{\sigma,\tau, \nu= \alpha, \beta}\hat{c}^{\dagger}_{p_1 \sigma}\hat{c}^{\dagger}_{p_2 \tau}\hat{c}^{\dagger}_{p_3 \nu}\hat{c}_{q_3 \nu}\hat{c}_{q_2 \tau}\hat{c}_{q_1 \sigma},
\end{aligned}
    \label{eq1}
\end{equation}
where $\hat{c}^{\dagger}_{p_1 \sigma}$ ($\hat{c}_{p_1 \sigma}$) represent the creation (annihilation) operator of the spatial orbital $p$ with spin $\sigma$. With these definitions, the reduced density matrices (RDMs) are directly defined as:
\begin{equation}
\begin{aligned}
    \Gamma^{p_1}_{q_1} &=\langle\Psi |\hat{E}^{p_1}_{q_1} |\Psi\rangle,\\[8pt]
    \Gamma^{p_1 p_2}_{q_1 q_2}&=\langle\Psi|  \hat{E}^{p_1 p_2}_{q_1 q_2}|\Psi\rangle,\\[8pt]
    \Gamma^{p_1 p_2 p_3}_{q_1 q_2 q_3}&=\langle\Psi| \hat{E}^{p_1 p_2 p_3}_{q_1 q_2 q_3} |\Psi\rangle.
\end{aligned}
    \label{eq2}
\end{equation}
Unlike their spin-orbital counterparts, the spin-free operators and RDMs lack anti-symmetries with respect to permutation of upper and lower indices. Working with spin-free operators improves memory usage and performance in the computational implementation. Moreover, since we seek to transform the Hamiltonian to seniority-zero form and we use a singlet reference (see Section.~\ref{sz-reference}), the spin-orbital and spin-free results agree.

\subsection{The SZ-QCT method}
\label{SZ-QCT}
In second quantization, the electronic Hamiltonian is
\begin{equation}
\begin{aligned}
\hat{H}= \sum_{p,q}h_{pq}\hat{E}^{p}_{q} +\frac{1}{2}\sum_{p,q,r,s}v_{pqrs}\hat{E}^{pq}_{rs},
\end{aligned}
    \label{eq3}
\end{equation}
where $h_{pq}$ and $v_{pqrs}$ are the one- and two-body electron integrals, and $\hat{E}^{p}_{q},\hat{E}^{pq}_{rs}$ are the previously defined spin free excitation operators over the general spatial orbitals $p,q,r,s$. Our method uses a unitary transformation
\begin{equation}
\begin{aligned}
\hat{\widetilde{H}}=e^{\hat{A}}\hat{H}e^{-\hat{A}} ,
\end{aligned}
    \label{eq4}
\end{equation}
to reduce the non-seniority-zero couplings of the final Hamiltonian $H_{SZ}$, where the generator of the transformation $\hat{A}$ is the anti-hermitian operator:
\begin{equation}
\begin{aligned}
\hat{A} =\sum_{p,q}a_{pq}\left(\hat{E}^{p}_{q} - \hat{E}^{q}_{p} \right) +\frac{1}{2}\sum_{p,q,r,s}a_{pqrs}\left(\hat{E}^{pq}_{rs} -\hat{E}^{rs}_{pq} \right),
\end{aligned}
    \label{eq5}
\end{equation}
with $a_{pq}$ and $a_{pqrs}$ are the generator one- and two-body amplitudes, and we assume that $a_{pqrs}$ is anti-symmetric with respect to interchange of $p$ and $q$ or $r$ and $s$. When all electrons are paired, most of the terms in the Hamiltonian drop out, giving the seniority-zero Hamiltonian, 

\begin{equation}
\begin{aligned}
\hat{H}_{SZ}=&2\sum_{p}h_{pp}\hat{N}_{p}
+ \sum_{p,q}v_{p,p,q,q}\,\hat{P}^{\dagger}_{p}\hat{P}_{q} \\[5pt]
&+ \sum_{p \neq q}\left(2v_{p,q,p,q} - v_{p,q,q,p}\right)\hat{N}_{p}\hat{N}_{q},    
\end{aligned}
\label{eq6}
\end{equation}
where $\hat{P}^{\dagger}_{p}=a^{\dagger}_{p\alpha}a^{\dagger}_{p\beta}$ and $\hat{P}_{p}=a_{p\beta}a_{p\alpha}$ are the pair creation and annihilation operators on spatial orbital $p$, and $\hat{N}_{p}=\hat{P}^{\dagger}_{p}\hat{P}_{p}$ is the pair-number operator. The first term counts the number of electron pairs in a given spatial orbital, the second term moves an electron pair from orbital $q$ to orbital $p$, and the last term
determines the number of electron pairs in two (spatial) orbitals. 

The idea behind SZ-QCT is to select the generator $\hat{A}$ that minimizes the non-seniority-zero elements of the transformed Hamiltonian, effectively mapping the system Hamiltonian into the seniority-zero space. We write the idea as the following optimization problem:
\begin{equation}
\begin{aligned}
&\min_{\hat{A}}\left(|| \hat{\widetilde{H}}_{\text{non-}SZ}||\right), \\[10pt]
&\hat{\widetilde{H}}_{\text{non-}SZ}= \hat{\widetilde{H}}-\hat{\widetilde{H}}_{SZ},
\label{eq7}
\end{aligned}
\end{equation}
where $\hat{\widetilde{H}}_{SZ}$ is the seniority-zero sector of the transformed Hamiltonian $\hat{\widetilde{H}}$.
\subsection{Reference wave function}
\label{sz-reference}
We choose to use an orbitally optimized doubly occupied configuration (OPT-DOCI) as our reference wave function. The cost of evaluating this wave function is the number of iterations times the square root of the FCI cost. Although this cost remains factorial and prohibitive for larger systems, many wave functions have been proposed in the literature that approximate DOCI with a much more affordable cost\cite{surjanIntroductionTheoryGeminals1999,rassolov2002geminal,surjan2012strongly,parr1956generalized,kutzelnigg1964direct,jeszenszki2015local,pernal2013equivalence,hurley1953molecular,parks1958theory,hunt1972self,hay1972generalized,small2014coupled,lawler2008symmetry,cullen1996generalized,moss1975generalized,dykstra1980perfect,carter1988correlation,hartke1992ab,doi:10.1021/ct300902c,PhysRevB.89.201106,10.1063/1.4880819,Boguslawski2015,JOHNSON2013101,moissetDensityMatricesSeniorityzero2022,faribaultReducedDensityMatrices2022,fecteauNearexactTreatmentSeniorityzero2022,fecteauReducedDensityMatrices2020,johnsonBivariationalPrincipleAntisymmetrized2022,johnsonRichardsonGaudinMeanfield2020,johnsonRichardsonGaudinStates2024,tecmerAssessingAccuracyNew2014a,henderson2025jordanwignertransformationdescriptionstrong}. Beyond capturing static correlation \cite{CaleroOsorio2025SZ}, DOCI and related seniority-zero wave functions have attractive features for use in multireference dynamic-correlation methods, such as the sparsity and block structure of their reduced density matrices (RDMs).\cite{weinholdReducedDensityMatrices1967,poelmansVariationalOptimizationSecondOrder2015a,head-marsdenPair2electronReduced2017,head-marsdenActiveSpacePairTwoElectron2020,fecteauReducedDensityMatrices2020,faribaultReducedDensityMatrices2022,moissetDensityMatricesSeniorityzero2022} Intuitively, this follows from the fact that the only non-zero elements of an $n$-body RDM are those that preserve the number of electron pairs (i.e., conserve seniority within the paired subspace). As a result, many index patterns of the RDMs are forbidden (e.g. $\Gamma^{p}_{q}=0$ for $p\neq q$), and the number of independent indices is reduced by a factor of two. A detailed presentation of the non-zero blocks of the RDMs is given in the Supplementary material. While our method is an alternative to other methods for adding dynamic correlation to a seniority-zero wavefunction ansatz,\cite{lawler2008symmetry,cullen1996generalized,moss1975generalized,carter1988correlation,huUnitaryBlockCorrelatedCoupled2025,liBlockcorrelatedCoupledCluster2004,robb1970generalized,robb1971generalized,robb1972generalized,kobayashiGeneralizedMollerPlessetPartitioning2010,limacher2014simple,alcobaHybridConfigurationInteraction2014a,vanraemdonckPolynomialScalingApproximations2015a,margocsyMultipleBondBreaking2018,zobokiLinearizedCoupledCluster2013,10.1063/1.4906607,garza2015range,Boguslawski2015,kimFlexibleAnsatzNbody2021,tecmer2022geminal,gaikwadCoupledClusterinspiredGeminal2024,miranda-quintanaFlexibleAnsatzNBody2024}  it is distinguished from alternative approaches in practice (e.g., it does not "freeze" the seniority-zero portion of the wavefunction, thereby allowing electron pairs to relax in the presence of dynamic correlation) and perspective (i.e., we strive to transform the Hamiltonian into a form where the seniority-zero treatment would be exact, rather than correct an approximate wavefunction \cite{10.1063/5.0309818,doi:10.1021/acs.jctc.5c01892}). 

\subsection{Commutator approximation and recursive expression}
\label{commutator_approx}
The evaluation of the unitary transformation of Eq.~\ref{eq4} is carried out using the Baker-Campbell-Hausdorff (BCH) expansion. However, given the unitary character of the transformation, the expansion does not truncate. Moreover, each additional term in the expansion increases the operator rank. For that reason, we use the recursive commutator approximation (RCA) from canonical transformation (CT) theory. In this approximation, the higher-order operators arising from the commutator $[\hat{X},\hat{Y}]$ are expanded in a series of products of lower-order operators and RDMs. The operators $\hat{X}$ and $\hat{Y}$ can represent any operators, e.g., the Hamiltonian, the generator, or a commutator itself. In our previous work, SZ-LCT, we used the single-commutator approximation from CT theory to evaluate the BCH expansion, where all single commutators $[\hat{H},\hat{A}]$ are approximated by at most two-body operators, $[\hat{H},\hat{A}]_{1,2}$. In the present work, following quadratic-CT,\cite{10.1063/1.3086932} we extend SZ-LCT by allowing double commutators to contribute to the BCH expansion as well. Thus, instead of applying the operator decomposition process to all commutators, we delay this operation until the double commutators:
\begin{equation}
    \underset{\text{SZ-LCT}}{\underbrace{[[\hat{H},\hat{A}]_{1,2},\hat{A}]_{1,2}}} \longrightarrow \underset{\text{SZ-QCT}}{\underbrace{[[\hat{H},\hat{A}],\hat{A}]_{1,2}}}
    \label{eq8}
\end{equation}
By doing this in SZ-QCT, we consider effective contributions from the four-body operators arising from the double commutator (see Section~\ref{op-decomp}). Thus, the approximate BCH expansion used for the unitary transformation in SZ-QCT is:
\begin{equation}
\begin{aligned}
\hat{\widetilde{H}} = \hat{H} + \left[\hat{H},\hat{A}\right]_{1,2} + \frac{1}{2!} \left[\left[\hat{H},\hat{A}\right],\hat{A}\right]_{1,2}  + \frac{1}{3!} \left[\left[\left[\hat{H},\hat{A}\right],\hat{A}\right]_{1,2}, \hat{A}\right]_{1,2} + \frac{1}{4!} \left[\left[\left[\left[\hat{H},\hat{A}\right],\hat{A}\right]_{1,2}, \hat{A}\right],\hat{A}\right]_{1,2} +...,
\end{aligned}
    \label{eq9}
\end{equation}
Given that the expansion contains both single- and double-commutator approximations, the recursive formula is not the same as in SZ-LCT. In particular, the $n$th term depends on either the $(n-1)$th or the $(n-2)$th term, depending on whether $n$ is odd or even.

To see this more clearly, consider the first terms of the previous expansion. Let us define $\bar{H}_0 = H$, $\bar{H}_1 = [H,A]_{1,2}$, and so on. Then, the first five terms of the expansion are
\begin{equation}
\begin{aligned}
\bar{H}_0 &= H,\\[5pt]
\bar{H}_1 &= [H,A]_{1,2} = [\bar{H}_0,A]_{1,2}~,\\[5pt]
\bar{H}_2 &= \frac{1}{2}[[H,A],A]_{1,2}= \frac{1}{2}[[\bar{H}_0 ,A],A]_{1,2}~,\\[5pt]
\bar{H}_3 &= \frac{1}{3!}[[[H,A],A]_{1,2},A]_{1,2}= \frac{1}{3}[\bar{H}_2 ,A]_{1,2}~,\\[5pt]
\bar{H}_4 &= \frac{1}{4!}[[[[H,A],A]_{1,2},A],A]_{1,2}= \frac{1}{4(3)}[[\bar{H}_2 ,A],A]_{1,2}~,\\[5pt]
\end{aligned}
    \label{eq10}
\end{equation}
which leads us to the following recursive expression
\begin{equation}
\label{eq11}
\bar{H}_n = 
\begin{cases}
\dfrac{1}{n(n-1)} \bigl[\!\bigl[\bar{H}_{n-2},A\bigr],A\bigr]_{1,2}, & \text{if $n$ is even},\\[8pt]
\dfrac{1}{n} \bigl[\bar{H}_{n-1},A\bigr]_{1,2}, & \text{if $n$ is odd}.
\end{cases}
\end{equation}
Therefore, even terms of the BCH expansion are computed using the approximation of the double-commutator while odd terms are computed using the single-commutator approximation.
\subsection{Operator decompositionn}
\label{op-decomp}
The single- and double-commutator approximation is an operation first developed in CT theory that allows high-body operators (i.e., three- and four-body operators) arising from the commutator $[\hat{X},\hat{Y}]$ to be written in terms of one- and two-body operators and RDMs. The expressions for the RCA are obtained using operator decomposition (OD), a method based on the generalized normal ordering (GNO) of Mukherjee and Kutzelnigg \cite{mukherjee1997normal,kutzelnigg1997normal,kutzelnigg2007generalized,kong2010algebraic} 
that extends the standard particle-hole normal ordering to multireference wave functions. The OD process uses the expressions for the spin-free GNO excitation operators and the core assumption that the reference wave function provides a sufficiently accurate description of the system such that the three- and four-body GNO excitation operators vanish. In this way, the three- and four-body spin-free excitation operators (not normal ordered) can be expanded in series of one- and two-body excitations and RDMs. The derivation of the three-body operator decomposition case has been presented elsewhere \cite{doi:10.1021/acs.jctc.5c01892,10.1063/1.3086932}. Here, we present the expression for the four-body spin-free operator decomposition, an expression that, to the best of our knowledge, has not been presented before.

The spin-free GNO excitation operators are:
\begin{equation}
\begin{aligned}
    \tilde{E}^{p_1}_{q_1}=&{E}^{p_1}_{q_1} - {\Gamma}^{p_1}_{q_1},\\[8pt]
    \tilde{E}^{p_1 p_2}_{q_1 q_2}=&~ {E}^{p_1 p_2}_{q_1 q_2} - \sum\left(-\frac{1}{2}\right)^{x} {\Gamma}^{p_1}_{q_1}\tilde{E}^{p_2}_{q_2} -{\Gamma}^{p_1 p_2}_{q_1 q_2},\\[8pt]
    \tilde{E}^{p_1 p_2 p_3}_{q_1 q_2 q_3}  = &~{E}^{p_1 p_2 p_3}_{q_1 q_2 q_3} - \sum\left(-\frac{1}{2}\right)^{x}\Big[{\Gamma}^{p_1}_{q_1}\tilde{E}^{p_2 p_3}_{q_2 q_3} \\[5pt] &- {\Gamma}^{p_1 p_2}_{q_1 q_2} \tilde{E}^{p_3}_{q_3}\Big] - \Gamma^{p_1 p_2 p_3}_{q_1 q_2 q_3},\\[8pt]
    \tilde{E}^{p_1 p_2 p_3 p_4}_{q_1 q_2 q_3 q_4}  &= {E}^{p_1 p_2 p_3 p_4}_{q_1 q_2 q_3 q_4} - \sum\left(-\frac{1}{2}\right)^{x}\Big[{\Gamma}^{p_1}_{q_1}\tilde{E}^{p_2 p_3 p_4}_{q_2 q_3 q_4}  \\[5pt]
    &\quad  - {\Gamma}^{p_1 p_2}_{q_1 q_2} \tilde{E}^{p_3 p_4}_{q_3 q_4} - {\Gamma}^{p_1 p_2 p_3}_{q_1 q_2 q_3} \tilde{E}^{p_4}_{q_4} \Big] - \Gamma^{p_1 p_2 p_3 p_4}_{q_1 q_2 q_3 q_4},
\end{aligned}
    \label{eq12}
\end{equation}
where the notation $\sum\left(-\frac{1}{2}\right)^xA^{p_1 p_2...p_k}_{q_1 q_2 .... q_k}B^{p_{k+1} p_{k+2}..}_{q_{k+1} q_{k+2} ....}$  means that there is one term for each partition of the indices ${p_i},{q_i}$ among the objects $A$ and $B$, where the ${p_i}$ are kept on top and the ${q_i}$ are kept on bottom, and a $\left(-\frac{1}{2}\right)$ factor is applied for each permutation that changes the original pairing. 

Assuming that $\tilde{E}^{p_1 p_2 p_3 p_4}_{q_1 q_2 q_3 q_4},\tilde{E}^{p_1 p_2 p_3}_{q_1 q_2 q_3} \rightarrow 0$ we substitute the expressions of $\tilde{E}^{p_1}_{q_1}$ and $\tilde{E}^{p_1 p_2}_{q_1 q_2}$ into the last equality of Eq.~\ref{eq12} and obtain Eq~\ref{eq13}. 
\begin{equation}
\begin{aligned}
E^{p_1 p_2 p_3 p_4}_{q_1 q_2 q_3 q_4}
&=
\sum(-\tfrac12)^x \gamma^{p_1 p_2}_{q_1 q_2} E^{p_3 p_4}_{q_3 q_4}
+
\sum(-\tfrac12)^x \gamma^{p_1 p_2 p_3}_{q_1 q_2 q_3} E^{p_4}_{q_4}
-2\sum(-\tfrac12)^x \gamma^{p_1 p_2}_{q_1 q_2}\gamma^{p_3}_{q_3}E^{p_4}_{q_4}
+
\sum(-\tfrac12)^x \gamma^{p_1 p_2}_{q_1 q_2}\gamma^{p_3}_{q_4}E^{p_4}_{q_3}
\\[10pt]
&\quad
+2\sum(-\tfrac12)^x \gamma^{p_1 p_2}_{q_1 q_2}\gamma^{p_3}_{q_3}\gamma^{p_4}_{q_4}
-
\sum(-\tfrac12)^x \gamma^{p_1 p_2}_{q_1 q_2}\gamma^{p_3}_{q_4}\gamma^{p_4}_{q_3}
-\sum(-\tfrac12)^x \gamma^{p_1 p_2 p_3}_{q_1 q_2 q_3}\gamma^{p_4}_{q_4}
+
\gamma^{p_1 p_2 p_3 p_4}_{q_1 q_2 q_3 q_4}
\end{aligned}
\label{eq13}
\end{equation}
This gives the decomposition of the four-body spin-free excitation operator and is used in the double-commutator approximation. Note that the previous expression includes the 4RDM. If this quantity is too expensive to compute for the chosen reference wave function, cumulant decompositions can be used \cite{Kutzelnigg10022010,10.1063/1.3256237,PhysRevA.47.979}, expressing the 4RDM in terms of one-, two-, and three-electron RDMs, or even only one- and two-electron RDMs. We implemented the decomposition of the four-body spin-free excitation operator in the $\texttt{sqa}$ software package \cite{10.1063/1.3086932}. 

\section{Implementation}
\label{implementation}
The reference wave function is obtained using a development version of \texttt{PyCI} together with \texttt{HORTON3} to perform an orbital-optimized DOCI calculation \cite{chanTaleHORTONLessons2024,10.1063/5.0219010} using the one- and two-body molecular electron integrals \cite{10.1063/5.0006074,sunLibcintEfficientGeneral2015,kimGBasisPythonLibrary2024}.
The symbolic expressions for the spin-adapted single- and double-commutator approximations ($[\hat{H},\hat{A}]_{1,2}$, $[[\hat{H},\hat{A}],\hat{A}]_{1,2}$) are obtained from an extended version of the \texttt{sqa} software package. The expressions were parsed and evaluated numerically using \texttt{NumPy} \texttt{einsum} and \texttt{opt\_einsum} \cite{daniel2018opt}. The minimization of Eq.~\ref{eq7} is performed by passing to a \texttt{SciPy} minimizer a function that computes the norm of the non-seniority-zero elements of the transformed Hamiltonian. Since the commutator approximation is valid only for small generators $\hat{A}$, we need to constrain the norm of $\hat{A}$, so we use the \textit{SLSQP} and \textit{trust-constr} algorithms. At convergence, the minimizer selects the generator $\hat{A}$ that minimizes the non-seniority-zero elements of the transformed Hamiltonian, effectively mapping the Hamiltonian into the seniority-zero subspace.

\subsection{Cost and performance}
The two central operations of the method are the commutator decomposition and the analytical gradient of the transformation. The naive cost of both $[\hat{H},\hat{A}]_{1,2}$ and $[[\hat{H},\hat{A}],\hat{A}]_{1,2}$ is $\mathcal{O}(N^7)$, where $N$ is the number of spatial orbitals. Although the scaling of both decompositions is the same, they differ in the prefactor, with the double commutator having a considerably larger number of unique terms (cf. Table~\ref{scaling_methods}), which makes it slower in practice. To reduce the scaling, we exploited the seniority-zero structure of the RDMs by contracting only over the nonzero index-pattern combinations, thus reducing the scaling to $\mathcal{O}(N^5)$. Furthermore, we recast the RDMs into a set of smaller tensors to avoid operating with tensors containing more than four indices. A detailed explanation of this algorithm can be found elsewhere \cite{doi:10.1021/acs.jctc.5c01892}.

On the other hand, the analytical gradient scales as $\mathcal{O}\left(\frac{N^2 (N-1)^2}{4}\right)$, which for fewer than approximately 150 spatial orbitals is effectively $\mathcal{O}(N^3)$. Since we parallelized the gradient evaluation such that multiple elements can be computed concurrently, the gradient cost becomes $\mathcal{O}\left(\frac{N^3}{n_c}\right)$ for small- and moderate-sized systems. With these optimizations, the effective overall scaling of the method is reduced to $\mathcal{O}\left(\frac{N^8}{n_c}\right)$. This is the same scaling as SZ-LCT and LT-SZCT.
\begin{table}[h!]
\caption{Number of unique terms and effective computational scaling for SZ-LCT, SZ-QCT, and LT-SZCT. The number of unique terms is given for the corresponding commutator approximation, while the scaling refers to the overall computational cost of each method.}
\label{scaling_methods}
\small
\centering
\renewcommand{\arraystretch}{1.15}
\begin{tabularx}{\columnwidth}{@{}l >{\centering\arraybackslash}X c c@{}}
\toprule
Method & Commutator & Unique Terms & Scaling \\
\addlinespace[2pt]
\midrule
\addlinespace[3pt]

SZ-LCT
& $[\hat{H},\hat{A}]_{1,2}$
& 68
& $\mathcal{O}\left(\frac{N^8}{n_c}\right)$ \\

\addlinespace[4pt]

SZ-QCT
&
\begin{tabular}[c]{@{}c@{}}
$[\hat{H},\hat{A}]_{1,2}$ \\
$[[\hat{H},\hat{A}],\hat{A}]_{1,2}$
\end{tabular}
& 7816
& $\mathcal{O}\left(\frac{N^8}{n_c}\right)$ \\

\addlinespace[4pt]

LT-SZCT
&
\begin{tabular}[c]{@{}c@{}}
$[\hat{H},\hat{A}]$ \\
$[[\hat{H},\hat{A}],\hat{A}],\ [\hat{H},\hat{A}]_{1,2}$
\end{tabular}
& 128
& $\mathcal{O}\left(\frac{N^8}{n_c}\right)$ \\

\addlinespace[3pt]
\bottomrule
\end{tabularx}
\end{table}

\section{Results}
\label{results}
\subsection{\ce{H6}}

\begin{figure}[!t]
\centering
\includegraphics[width=0.7\linewidth]{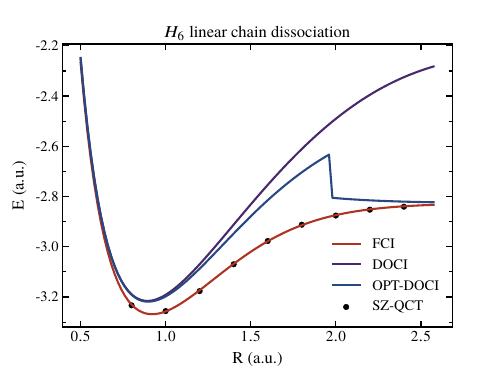}
\caption{\label{fig1} Dissociation curve for the linear \ce{H6} chain in the STO-6G basis set as a function of nearest-neighbor distance.}
\end{figure}

We start with the symmetric stretch of the linear \ce{H6} molecule in the STO-6G basis set. As mentioned before, orbital-optimized DOCI (OPT-DOCI) is the reference wave function for our tests. We calculate the ground state energy with FCI, DOCI, OPT-DOCI and SZ-QCT method and plot them in Fig.~\ref{fig1}.

\begin{figure}[!t]
\centering
\includegraphics[width=0.7\linewidth]{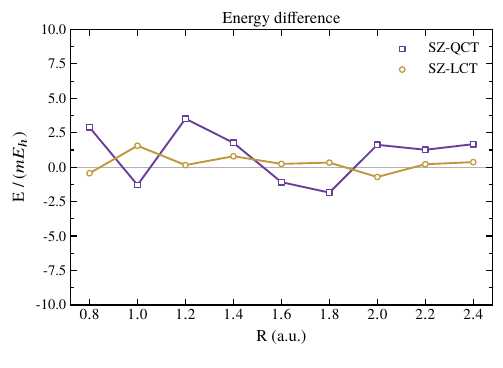}
\caption{\label{fig2} Energy deviations relative to FCI for SZ-QCT and SZ-LCT, in $mE_h$, for the linear \ce{H6} dissociation in the STO-6G basis set}
\end{figure}

To compare the accuracy of SZ-QCT and the previously proposed SZ-LCT, we plotted both energy differences with respect to FCI in Fig.~\ref{fig2}. From this figure we first notice that SZ-QCT shows good accuracy, with a maximum energy error of $3.52~mE_h$ and average error of $1.89~mE_h$. However, contrary to our expectations, SZ-QCT proves to be less accurate than SZ-LCT.

Furthermore, the method inherits the erratic jumps in the energy difference previously discussed for SZ-LCT, which we attributed to convergence to nearly degenerate local minima with nearly the same value of the objective function. Although SZ-QCT is formulated to relax the constraint over the transformation generator, it also significantly increases the complexity of the objective function by adding $\sim$ seven thousand extra unique terms to the commutator approximation. This extra complexity affects the structure of the variational space, so additional local minima emerge, some of which are probably significantly further away from the exact ground-state energy.

It is important to note that the SZ-QCT results remain good even when the OPT-DOCI reference has converged to a spurious local minima ($\sim 2.0$ a.u.). It is reassuring that the SZ-QCT corrections reduces the sensitivity of OPT-DOCI to being trapped in local minima during the orbital optimization procedure.

Preliminary tests on SZ-QCT showed a deterioration in the method's accuracy when non-optimized DOCI orbitals were used, particularly at stretched geometries near dissociation. This occurs because, in that regime, the double-commutator approximation breaks down when the reference wave function no longer provides an adequate approximation to the exact wave function. This behavior was also observed for SZ-LCT; however, the decrease in accuracy was less severe for SZ-QCT than for SZ-LCT, since the former is mathematically designed to allow larger generators in the transformation.

\subsection{\ce{BeH2}}
\begin{figure}[!t]
\centering
\includegraphics[width=0.7\linewidth]{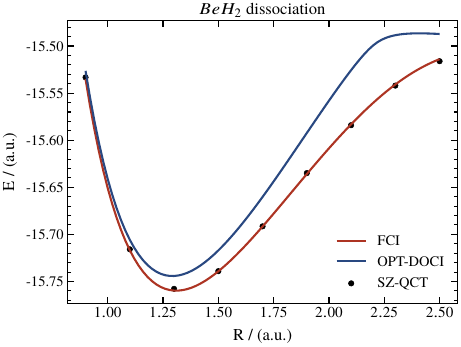}
\caption{\label{fig3} Dissociation energy for the linear \ce{BeH2} molecule in the STO-6G basis set as a function of nearest-neighbor distance.}
\end{figure}

\begin{figure}[!t]
\centering
\includegraphics[width=0.7\linewidth]{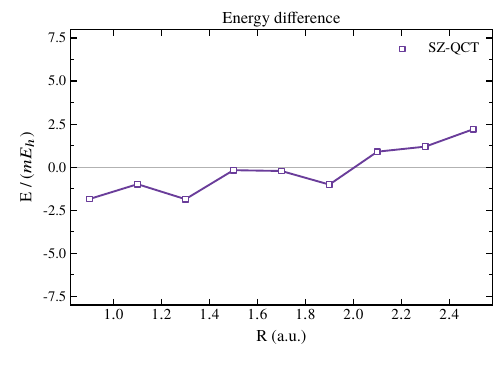}
\caption{\label{fig4} The energy difference between the SZ-QCT and FCI, in $mE_h$, for the dissociation of the linear \ce{BeH2} molecule in the STO-6G basis set.}
\end{figure}

Next, we consider the symmetric stretching of linear \ce{BeH2} in the STO-6G basis set. This molecule is particularly well described by OPT-DOCI from equilibrium to moderately stretched geometries (e.g., Fig.~\ref{fig3}), with energy differences of $\sim 10~mE_h$ for $R \leq 1.5$ a.u. At larger distances, closer to dissociation, the reference deviates from the exact result, with energy differences of a few tens of milliHartree. With this in mind, we expect SZ-QCT to deliver the most accurate results for small to intermediate internuclear distances and to deviate more strongly near dissociation, which is precisely the behavior observed in Fig.~\ref{fig4}.
For near-equilibrium distances, SZ-QCT shows the best results, with sub-millihartree energy differences. As expected, the largest energy error is found at $R = 2.6$ a.u. SZ-QCT delivers good accuracy; however, its computational time is significantly higher than that of SZ-LCT or LT-SZCT. This increased cost arises from the large number of unique terms present in the double-commutator approximation $[[\hat{H},\hat{A}],\hat{A}]_{1,2}$. For that reason, strategies that factorize the tensor contractions to reduce the overall scaling could be helpful, although they may not be sufficient.

Another possible explanation for why SZ-QCT is not more accurate than SZ-LCT is that, for the systems considered here, including approximate four-body contributions in the transformed Hamiltonian may not be physically necessary. As discussed above, the core assumption behind the RCA in CT theory and SZ-LCT is that the reference wave function is sufficiently close to the exact wave function such that the transformed Hamiltonian can be described accurately at the one- and two-body level. When this assumption already holds, retaining higher-body contributions may not improve the energy prediction, but instead may overcomplicate the optimization landscape and drive the minimization toward less favorable local minima. For that reason, in the next test case we consider a more challenging system for OPT-DOCI.

\subsection{\ce{N2}}
\begin{figure}[!h]
\centering
\includegraphics[width=0.7\linewidth]{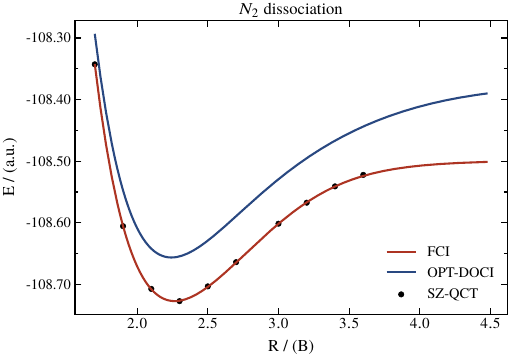}
\caption{\label{fig5}  \ce{N2} dissociation energy curve in STO-6G basis set.}
\end{figure}

\begin{figure}[!h]
\centering
\includegraphics[width=0.7\linewidth]{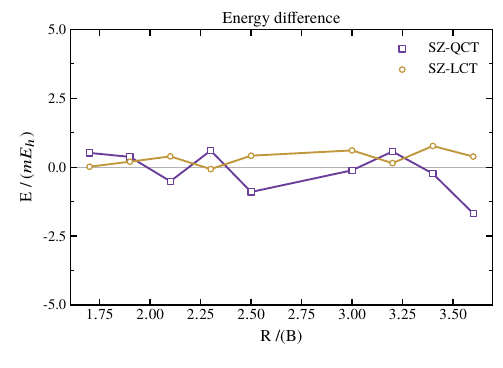}
\caption{\label{fig6}Energy deviations relative to FCI of the SZ-QCT and SZ-LCT methods, in $mE_h$, for \ce{N2} dissociation in STO-6G basis set.}
\end{figure}

Lastly, we consider the \ce{N2} molecule dissociation in the STO-6G basis set. It is well known that multiple bond dissociation is a challenging case for OPT-DOCI (e.g. see Figure.~\ref{fig5}). So, to accurately include the missing correlation, bigger generators are needed, mostly in the near-dissociation geometries.
Energy differences relative to FCI for the SZ-QCT and SZ-LCT method are shown in Figure.~\ref{fig6}. With most of the energy errors within sub-milli Hartree, \ce{N2} proves to be the best scenario for the application of SZ-QCT method. Although SZ-QCT is still marginally less accurate than SZ-LCT, the accuracy difference is no longer of order of magnitude. Moreover, for stretched geometries, where the flexibility of the SZ-QCT generators is most needed, we see two points, $R=2.7$a.u and $R=3.4$a.u, where the accuracy of SZ-QCT overpass that of SZ-LCT. Nevertheless, for the most stretched point we see SZ-LCT's accuracy is better, showing that although SZ-QCT is well suited for physical cases where the OPT-DOCI deviates from the true wave function, the problems associated with the convergence to local minima persist.

\section{Conclusions}
\label{conclusions}
We present a method for solving the Schrödinger equation using Hamiltonian transformations. Specifically, We use a unitary transformation to downfold the Hamiltonian into the seniority-zero subspace, thereby reducing the computational cost of the final solution. The transformation is evaluated using the BCH expansion, for which we employed strategies first introduced in canonical transformation (CT) theory to evaluate the commutators recursively (RCA). We extended our previously proposed seniority-zero linear canonical transformation theory (SZ-LCT) \cite{10.1063/5.0309818} to partially retain approximate four-body contributions through the double-commutator approximation $[[\hat{H},\hat{A}],\hat{A}]_{1,2}$. This approximation uses the four-body spin-free operator decomposition expression, an expression that, to the best of our knowledge, had not been previously presented and/or tested.

The results show good accuracy, with most points lying within chemical accuracy. Like SZ-LCT, SZ-QCT inherits the convergence-to-local-minima problem, which manifests as discontinuities or jumps in the energy deviations relative to FCI. Nonetheless, these jumps are generally larger for SZ-QCT. We attribute this to the more complicated structure of the objective function in this method, which may make the minimization process more prone to (false) convergence to local minima.

SZ-QCT showed its best performance for the \ce{N2} molecule, particularly at stretched geometries, with sub-millihartree accuracy and surpassing SZ-LCT at a few points. Since \ce{N2} is the case where the orbital-optimized DOCI (OPT-DOCI) reference is least accurate, SZ-QCT appears to be better suited for cases in which larger generators are needed to recover the missing residual correlation. In regimes where OPT-DOCI is sufficiently close to the exact wave function, the SZ-QCT formulation appears to exacerbate the optimization problem, leading to a deterioration in the quality of the found solutions.

A comparison with SZ-LCT was carried out for two of the three tested molecules. For \ce{H6}, where OPT-DOCI provides a good description of the exact wave function, SZ-LCT is one, and in some cases two, orders of magnitude more accurate than SZ-QCT. For \ce{N2}, both SZ-LCT and SZ-QCT show similar errors for most bond lengths. Even so, the overall performance of SZ-LCT remains better than that of SZ-QCT on average. Although both methods have the same formal scaling, the prefactor of SZ-QCT is significantly larger.

Based on these results, we conclude that the double-commutator extension included in SZ-QCT does not improve the accuracy of the method for most of the cases considered here, making SZ-LCT the more suitable choice for adding dynamic correlation in small- to medium-sized systems. 
This result is at first surprising, but it is consistent with the behavior reported for the quadratic extension of CT theory (QCT) \cite{10.1063/1.3086932}. In that work, QCT was generally less accurate than the linear version (LCT) when the reference wave function had multiconfigurational character. Improvements from QCT were mainly observed when a Hartree--Fock reference was used, which is analogous to our observation that SZ-QCT improves only in cases where larger generators are needed. 

\section{acknowledgement}
The authors acknowledge support from the Canada Research Chairs (CRC-2022-00196) and NSERC (Discovery RGPIN/06707-2024 and Alliance ALLRP/592521-2023). 

\section{Research Resources}
Computing resources were provided by the Digital Research Alliance of Canada.

\section{Data Availability}
The data that support the findings of this study are available from the corresponding author upon reasonable request. Further data supporting our findings are also available in the supplementary material.

\bibliography{cas-refs}

\end{document}